\pdfoutput=1
\documentclass[sigconf]{acmart}
\usepackage[utf8]{inputenc}
\usepackage{subfigure}
\usepackage{graphicx}
\usepackage{amsmath}

\usepackage{amsfonts}
\usepackage{natbib}

\usepackage{graphicx}
\usepackage{latexsym}
\usepackage{booktabs}
\usepackage{multirow}
\usepackage{caption}
\usepackage[noend]{algpseudocode}
\usepackage{bbm}
\usepackage{algorithmicx,algorithm}
\usepackage{threeparttable}

\AtBeginDocument{%
  \providecommand\BibTeX{{%
    \normalfont B\kern-0.5em{\scshape i\kern-0.25em b}\kern-0.8em\TeX}}}

\setcopyright{acmcopyright}
\copyrightyear{2018}
\acmYear{2018}
\acmDOI{10.1145/1122445.1122456}

\acmConference[Woodstock '18]{Woodstock '18: ACM Symposium on Neural
  Gaze Detection}{June 03--05, 2018}{Woodstock, NY}
\acmBooktitle{Woodstock '18: ACM Symposium on Neural Gaze Detection,
  June 03--05, 2018, Woodstock, NY}
\acmPrice{15.00}
\acmISBN{978-1-4503-9999-9/18/06}

\begin{document}

\title{Balancing Multi-level Interactions for Session-based Recommendation}


\author{Yujia Zheng}
\authornote{Both authors contribute equally to the paper, and the order is decided by a coin flip.}
\authornote{Corresponding author: zhengyujia@std.uestc.edu.cn
}
\affiliation{
\institution{University of Electronic Science and Technology of China}}
\author{Siyi Liu}
\authornotemark[1]
\affiliation{
\institution{University of Electronic Science and Technology of China}}
\author{Zailei Zhou}
\affiliation{
\institution{University of Electronic Science and Technology of China}}


\begin{abstract}
Predicting user actions based on anonymous sessions is a challenge to general recommendation systems because the lack of user profiles heavily limits data-driven models. Recently, session-based recommendation methods have achieved remarkable results in dealing with this task. However, the upper bound of performance can still be boosted through the innovative exploration of limited data. In this paper, we propose a novel method, namely \emph{Intra-and Inter-session Interaction-aware Graph-enhanced Network}, to take inter-session item-level interactions into account. Different from existing intra-session item-level interactions and session-level collaborative information, our introduced data represents complex item-level interactions between different sessions. For mining the new data without breaking the equilibrium of the model between different interactions, we construct an intra-session graph and an inter-session graph for the current session. The former focuses on item-level interactions within a single session and the latter models those between items among neighborhood sessions. Then different approaches are employed to encode the information of two graphs according to different structures, and the generated latent vectors are combined to balance the model across different scopes. Experiments on real-world datasets verify that our method outperforms other state-of-the-art methods.
\end{abstract}


\maketitle
\section{Introduction}
Session-based Recommendation System (SRS) has attracted much attention for its highly practical value, especially in some real-world scenarios that concentrated with multitudes of anonymous interactive data (e.g., social media, e-commerce and web search) \citep{ludewig2018evaluation}. Different from most of the other recommendation tasks that need explicit user preference profiles, SRS only relies on anonymous user action logs (e.g, clicks) in an ongoing session to predict the user's next action \citep{ludewig2018evaluation,quadrana2018sequence}.

Under these circumstances, several methods are proposed to tackle the SRS task. Markov Chains (MC) \citep{zimdars01,shani2005mdp,rendle2010factorizing} is a representation of traditional methods. It predicts the user’s next action based on the previous one thus introduces sequentiality into SRS. Recently, neural network-based methods have become popular due to their strong abilities to model sequential data (e.g., Recurrent neural networks (RNN)-based networks). For instance, experiments conducted on real-world datasets \citep{hidasi2015session} show that Gated Recurrent Unit (GRU) significantly improves the performance of SRS compared with traditional methods. Based on \citep{hidasi2015session}, Tan et al. \citep{tan2016improved} improve the recommendation performance by applying data augmentation. After that, Li et al \citep{li2017neural} propose NARM to capture more representative features by using a global and local RNN structure. Similar to that, Liu et al. \citep{liu2018stamp} propose STAMP to model general and current interests using a novel attention mechanism. Moreover, as geometric deep learning methods (e.g., Graph Neural Networks (GNN)) have achieved state-of-the-art performance in various tasks, it is also applied in SRS after modeling sessions into graph-structured data \citep{wu2019srgnn}. 

Despite the surprising performance of deep learning methods, neighborhood-based methods can still provide competitive results. Traditional Item-based KNN (Item-KNN) \citep{sarwar2001item} considers the similarities between the last item in a current session and other items and recommends the most similar items to users. More recent methods, such as Session-based KNN (SKNN), consider each session as a whole and use the similarity between sessions to make recommendations \citep{ludewig2018evaluation,jannach2017recurrent,bonnin2014sknn}. Then, based on SKNN, KNN-RNN \citep{jannach2017recurrent} widens the application area by integrating GRU4REC to model the sequentiality. Most recently, CSRM \citep{wang2019CSRM} achieves state-of-the-art performance by applying Parallel Memory Modules on NARM to incorporate collaborative information. It calculates the similarities between the current and other session representations from the external memory module to extract collaborative information.

However, all existing methods are deficient in exploiting the depth or width of session data, or both. RNN-based methods \citep{hidasi2015session,li2017neural,liu2018stamp,jannach2017recurrent,wang2019CSRM}, including those combined with neighborhood-based methods \citep{jannach2017recurrent,wang2019CSRM}, only support unidirectional interactions between consecutive items and neglect those among other contextual items in the same session. SR-GNN \citep{wu2019srgnn} applies GNN to overcome this limitation but only within a small scope of a single session, which is a common deficiency in almost all purely neural network-based methods. Because there is no side-information (e.g., user profiles) in SRS, being poor at mining in a wider scope consisting of multiple sessions leads to a lack of collaborative information that limits the upper bound of model performance. Item-KNN-based methods do consider a large set of items that spread across multiple sessions, but they only treat items as independent elements, ignoring the integrity and sequentiality of sessions. While SKNN considers the integrity by measuring similarities between different sessions and its improved version KNN-RNN integrates GRU4REC to extract intra-session item-level interactions, interactions of items among different sessions are ignored by treating sessions as the minimum granularity. CSRM only introduces an end-to-end neural network method to push performance higher, but it does not address the aforementioned issue of coarse granularity and still suffers from the lack of contextual item-level interactions as RNN-based methods.

\begin{figure}[t]
    \centering
    \includegraphics[width=0.45\textwidth]{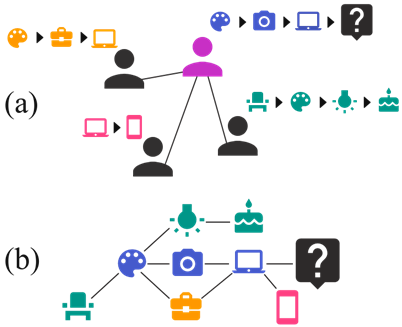}
    \caption{Here are three kinds of interactions. (a): Interactions between four users are Inter-session Session-level Interactions, and those of items within a single user's session are Intra-session Item-level Interactions. (b): Inter-session Item-level Interactions.}
    \label{figintro}
\end{figure}

Thus, we propose a novel method, namely \emph{Intra-and Inter-Session Interaction-aware Graph-enhanced Network} (I3GN), to reclaim the virgin land: inter-session item-level interactions. Figure \ref{figintro} visualizes the difference between the newly introduced interactions and others. The motivation of introducing that is to naturally model interactions within related items. For example, in Figure \ref{figintro}, a computer, which is in the current session, is linked with a camera, phone, and briefcase in different sessions. Therefore, we could assume that those items have certain commonalities, and each of these may have a relatively significant relevance to the computer. Other items in sessions of those related items, albeit to a lesser extent, may have similar effects. Previous methods, such as CSRM, unfortunately, mix related items with other relatively unimportant items together when they roughly divide them into different session representations, which are called inter-session session-level interactions (Figure \ref{figintro}a). However, our introduced inter-session item-level interactions explicitly highlight the importance of related items by considering all interactions among them (Figure \ref{figintro}b). Apparently, only focusing on that is not enough to make a strong recommendation. Therefore, we propose \emph{Session Merging Module} (SMM) to construct intra-and inter-session graphs for intra-and inter-session item-level interactions respectively. Because those related items share a certain commonality but in different relative positions for the computer, we use undirected edges to build the inter-session graph. As sequentialiy matters a lot when it comes to a single current session without any collaborative information, we choose the directed graph as the intra-session graph. Then \emph{Cross-Scope Encoder} (CSE) can be applied in these graphs to model complex items-level interactions across different scopes, using GNN's powerful ability of extracting rich contextual interactions by nodes propagation. In this way, not only contextual items in current sessions can be well represented but also different items in the inter-session graph can be given different weights based on their distance to the target item node. Because of the structural difference, we use different approaches to encode each kind of graph. Finally, for the current session, we combine the latent vectors of its two kinds of graphs and generate its representation vector for prediction.

The main contributions of our work are summarized as follows:
\begin{itemize}
    \item {We propose a novel graph-based model I3GN. To the best of our knowledge, it is the first method to integrate inter-session item-level interactions in SRS.}
    \item {We introduce a \emph{SMM} to model a current session into an intra-session graph and an inter-session graph, which lays a solid foundation for the extraction of complex item-level interactions.}  
    \item {We design a \emph{CSE} to balance the model in different kinds of session data. Inspired by SR-GNN, CSE introduces a wider scope of item-level interactions to boost the performance upper bound.}
    \item {We evaluate our model in two real-world datasets. Our extensive experiments show that I3GN outperforms the state-of-art methods.}
\end{itemize}

\section{Related Work}
\noindent In this section, we introduce some related works in Session-based Recommendation, Collaborative Filtering, and Graph Embedding.

\subsection{Session-based Recommendation}

\noindent Traditional methods for session-based recommendation are mainly based on Markov Chains (MC) \citep{zimdars01,Mobasher02,shani2005mdp,rendle2010factorizing}, which introduces the sequentiality in SRS by predicting the user's next action based on the last action. Zimdars et al. \citep{zimdars01} apply probabilistic decision-tree models to study the way to extract the sequentiality. Mobasher et al. \citep{Mobasher02} choose the contiguous sequential patterns for SRS after studying the effect of different patterns. Shani et al. \citep{shani2005mdp} employ the Markov Decision Processes that consider the long-term effect and the expected value of each recommendation. However, MC-based methods lose a balance between user's general preference and sequential behavior, for they seldom consider sequentiality between items that are not consecutively adjacent in the same session. To achieve that balance, Rendle et al. \citep{rendle2010factorizing} propose a hybrid method taking account of the combination of Matrix Factorization and MC, namely FPMC.

Like most other fields of recommendation, deep learning methods frequently appear in recent SRS models and obtain new state-of-the-art performance in terms of accuracy, especially RNN-based methods \citep{hidasi2015session,tan2016improved,li2017neural,liu2018stamp}. Hidasi et al. \citep{hidasi2015session} employ RNN with the Gated Recurrent Unit (GRU) into SRS and outperform traditional methods. Tan et al. \citep{tan2016improved} further improve it by introducing data augmentation, distillation integrating privileged information, and a pre-training approach to account for temporal shifts in the data distribution. Later, attention mechanism is applied by an encoder-decoder recommendation method (NARM) to combine sequentiality and user's general preference \citep{li2017neural}. However, STAMP \citep{liu2018stamp} also adopts the concept combining general and current interest, but the difference is that STAMP explicitly models the current interest reflected by the last click to emphasize the importance of last click, while NARM considers them as equally important. Most recently, geometric deep learning has become popular in a variety of tasks. SR-GNN \citep{wu2019srgnn} transforms sessions into the graph-structured data and applies GNN based on that. The significant improvement in recommendation performance proves the potential of geometric deep learning in SRS, and the motivation behind this work is enlightening to our work. However, all aforementioned deep learning methods only consider intra-session item-level interactions, which limits the upper bound of performance because of the lack of collaborative information.

Moreover, Collaborative Filtering (CF) idea-based methods are also popular in SRS. Unlike traditional user-based \citep{Resnick94,Hill95,Miller03,Jin04} or item-based \citep{Linden03,Billsus98,Pareek13,Sarwar00} CF models in other recommendation tasks, modifications need to be made for them to perform well in SRS. Simply using item neighborhood information \citep{sarwar2001item} cannot extract the integrity and sequentiality of items in the current session, which are extremely important for SRS application scenarios because of the lack of auxiliary data. Thus, SKNN \citep{bonnin2014sknn} is proposed to consider each session as a whole and its improved version KNN-RNN \citep{jannach2017recurrent} integrates GRU4REC to extract the sequentiality. Later, an end-to-end neural model (CSRM \citep{wang2019CSRM}) outperforms KNN-RNN with learnable latent session representations. The major difference between our method and theirs (KNN-RNN and CSRM) is that they only stay at the minimum granularity of session as the collaborative information, while we dig deeper and integrate inter-session item-level interactions into SRS. And the performance of CSRM is highly related to the quality of RNN-based encoder, which suffers from the lack of contextual item-level interactions that can be easily extracted by GNN in our method.
\begin{figure*}[ht]
    \centering
    \includegraphics[width=\textwidth]{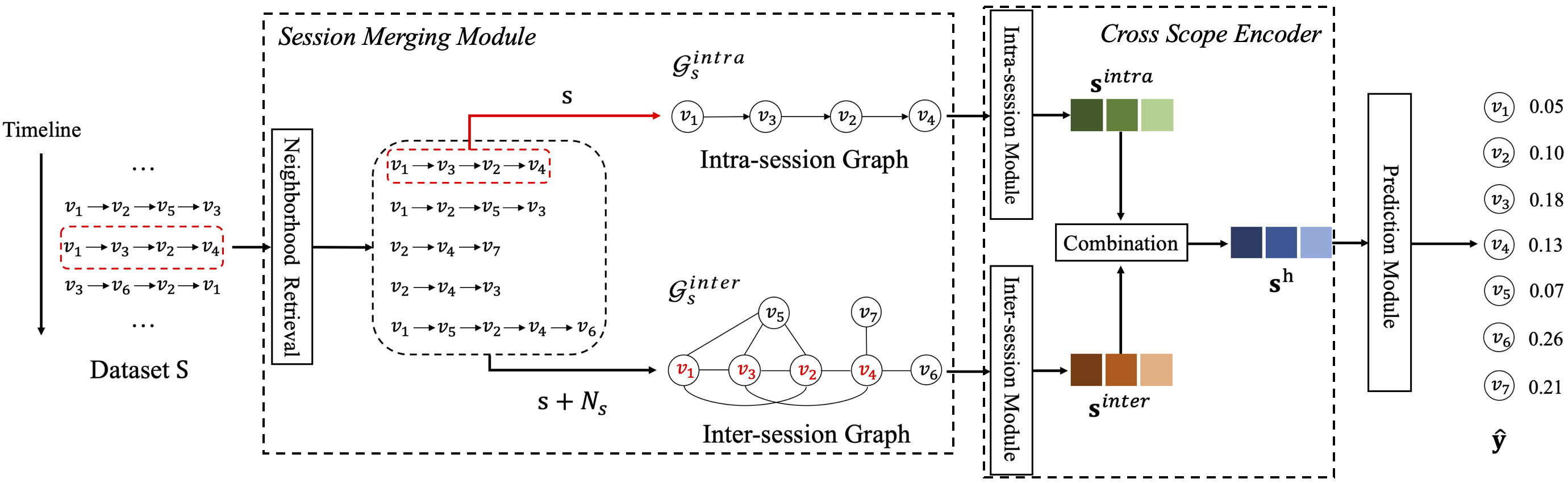}
    \caption{The framework of I3GN. 
    }
    \label{fig2}
\end{figure*}

\subsection{Graph Embedding}
\noindent The most important part of aforementioned deep learning-based methods is embedding, because generating more accurate and meaningful session embedding directly decides the performance. Thus, graph embedding becomes a critical component of the method when it comes to graph-structured data. However, traditional kernel-based methods (e.g., Weisfeiler-Lehman kernel \citep{shervashidze11}, Deep Graph Kernels \citep{Yanardag15}) focus more on the unsupervised tasks and have trouble scaling to large graphs, so we mainly introduce neural network-based graph embedding methods here. 

The concept of Graph Neural Networks (GNN) is first purposed by Gori et al. \citep{Gori05}, then developed and deepened by Scarselli et al. \citep{scarselli09} and Micheli et al. \citep{micheli09}. These early methods mainly generate representations of target nodes by using the recurrent neural unit to aggregate information of neighbor nodes. Inspired by the success of Convolutional Neural Network (CNN) in the image classification task, Bruna et al. \citep{Bruna14} propose the spectral Graph Convolutional Neural Network (GCN). Then Defferrard et al. propose a variant model by introducing fast localized spectral filtering \citep{Defferrard16}, and Kipf et al. improve it with a first-order approximation of spectral graph convolutions to motivate the choice of convolutional architecture \citep{Kipf16}. Moreover, Message Passing Neural Network (MPNN) generalizes these GCN-based methods and introduces a two-step framework: message passing and readout \citep{Gilmer17,Duvenaud15,Li16}. Gated Graph Neural Networks (GGNN) \citep{Li15} extends GNN to the sequential output, which is of great significance for sequential recommendations, such as SRS. However, as a large number of studies have shown that the attention mechanism improves the performance of deep learning-based methods in various tasks, it is therefore natural for researchers to import it on graphs \citep{Velickovic17,choi17}. Velickovic et al. \citep{Velickovic17} propose Graph Attention Network (GAT), which uses attention mechanisms to learn node embedding in a graph. By making the weights of nodes trainable, GAT can extract more information from the most critical part of the graph structure without a priori knowledge of structure, which is especially important for the scalability of graph embedding. Thus we apply it to our relatively large inter-session graphs.

\section{Model}
In this section, we introduce the proposed model. We start with a general introduction to the overall process of the model. Then, the internal structure of \emph{Intra-and Inter-Session Interaction-aware Graph-enhanced Network} (I3GN) is explained in detail.
\subsection{Notations}
The aim of SRS can be defined as using users’ current sequential session data to predict users’ next click items. Let $I = \{v_1, v_2, ... , v_{|I|}\}$ represents a set of unique items in all sessions. $s = \{v_1, v_2, ... , v_n\}$ represents an anonymous session which contains items ordered by timestamps. $S = \{s_1, s_2, ... , s_{|S|}\}$ denotes the whole sessions set. For each item in $I$, we embed it into a unified embedding space. Let $\mathbf{v}_i \in \mathbb{R}^{d}$ denotes the latent vector of corresponding item $v_i \in I$ and $ V = \{\mathbf{v}_1, \mathbf{v}_2, ... ,\mathbf{v}_{|I|}\}$ represents a set of all latent representations. Given a session $s$, the aim of our model is to predict user's possible next click item, i.e. the sequence label $v_{t+1}$. We generate probabilities $\widehat{\mathbf{y}}$ for all possible items based on input session $s$. Each element's value of vector $\widehat{\mathbf{y}}$ is the recommendation score of the corresponding item. The items with a top-$N$ recommendation score will be recommended as our model's output.

\subsection{Framework}
As illustrated in Figure \ref{fig2}, I3GN consists of following parts: \emph{Session Merging Module} (SMM), \emph{Cross-Scope Encoder} (CSE) and the final Prediction Module. Moreover, CSE can be divided into Intra-session Module and Inter-session Module according to encoding approaches of different graph structures. Specifically, for current session $s$, SMM uses a specific similarity to find the $k$ most similar sessions (neighbors) $N_s$ from all past sessions. Then intra-session graph $\mathcal{G}_s^{intra}$ is built based on $s$ and SMM uses $s$ and $N_s$ to construct inter-session graph $\mathcal{G}_s^{inter}$. According to $\mathcal{G}_s^{intra}$ and $\mathcal{G}_s^{inter}$, the representation of intra-and inter-session $\mathbf{s}^{intra}$ and $\mathbf{s}^{inter}$ are generated by Intra-and Inter-session Modules respectively. Then CSE combines them to generate the final session's representation $\mathbf{s}^h$. After that, based on $\mathbf{s}^h$, the Prediction Module produces the final output vector $\widehat{\mathbf{y}}$ as the recommendation scores for all possible items. Finally, the top-$N$ items in $\widehat{\mathbf{y}}$ are recommended.

\subsection{Session Merging Module}
To integrate inter-session item-level interactions, we need to model sessions into graph-structured data. 

Given a session $s$, the first step is to determine the neighbor set $N_s$ of the $k$ most similar past sessions in the training for graph-building. To achieve this, we construct the set of \emph{possible} neighbors $\mathcal{S}$ by creating the union of sessions in which the items of $s$ are contained. Recent study indicates that it is most effective to focus only on the most recent session when selecting neighbors \citep{ludewig2018evaluation}, so we create a subsample of $\mathcal{S}$ which contains $m$ most recent sessions denoted as $\mathcal{S}^{\prime} \subset \mathcal{S}$. In our method, we set $m$ to 1000 based on \citep{ludewig2018evaluation}.

After obtaining $\mathcal{S}^{\prime}$, we need to choose neighbors of $s$ from it. First, we compute the cosine similarities between $s$ and every other session $s_j \in \mathcal{S}^{\prime}$. Sessions $s$ and $s_j$ are encoded as binary vectors $\vec{s}, \vec{s_j} \in \mathbb{R}^{|I|}$, where if an item appears then the corresponding element in the vector is set to one, otherwise zero. Then we use cosine similarity to calculate the similarity between $\vec{s}$ and $\vec{s_j}$, which can be defined as:
\begin{equation} \label{eq1}
    \begin{split}
    sim(\vec{s},\vec{s_j}) = \frac{\vec{s} \cdot \vec{s_j}}{\sqrt{l(s) \cdot l(s_j)}}
    \end{split}
\end{equation}
where $l(s)$ and $l(s_j)$ represent the length of $s$ and $s_j$ respectively. 

For all sessions in $\mathcal{S}^{\prime}$, we first filter out sessions of which the similarity is lower than 0.5. Then the top-$k$ similar sessions are selected to create neighborhood set $N_s$. 

The second step is to build the intra-and inter-session graphs for intra-and inter-session item-level interactions, respectively. For intra-session graph, we model the current session $s = \{v_1, v_2, ... , v_n\}$ as a directed graph $\mathcal{G}_{s}^{intra}=\left(\mathcal{V}_{s}^{intra}, \mathcal{E}_{s}^{intra}\right)$ by treating each item $v_n \in \mathcal{V}_{s}^{intra}$ in $s$ as a node and $(v_{i-1}, v_i) \in \mathcal{E}_{s}^{intra}$ as an edge, which represents a user clicking on item $v_{i-1}$ and then clicking on $v_{i}$ in $s$. The reason why we use the directed graph is because sequentiality matters a lot when dealing with a single session.

For inter-session graph, we model all sessions in $s$ and $N_s$ as a single undirected inter-session graph $\mathcal{G}_{s}^{inter}=\left(\mathcal{V}_{s}^{inter}, \mathcal{E}_{s}^{inter}\right)$. In $\mathcal{G}_{s}^{inter}$, each node represents a item $v_i \in \mathcal{V}_{s}^{inter}$ that appears in session $s$ or any neighbor session in $N_s$, and each edge $(v_{i-1}, v_i) \in \mathcal{E}_{s}^{inter}$ denotes a user clicking on item $s_{n-1}$ before or after $s_{n}$ in session $s$ or any neighbor sessions in $N_s$. The motivation of why we use undirected graph is that related items might be located in different relative positions for target items in the current session. 

The visualization of this process is shown in Figure \ref{fig2}.

\subsection{Intra-session Module}
After obtaining the intra-session graph $\mathcal{G}_{s}^{intra}$, Intra-session Module is used to extract intra-session item-level interactions in session $s$ through the following two processes: \emph{Node representation learning} and \emph{Intra-session representation generating}, which are demonstrated as below.


\begin{figure}[t]
    \centering
    \includegraphics[width=0.45\textwidth]{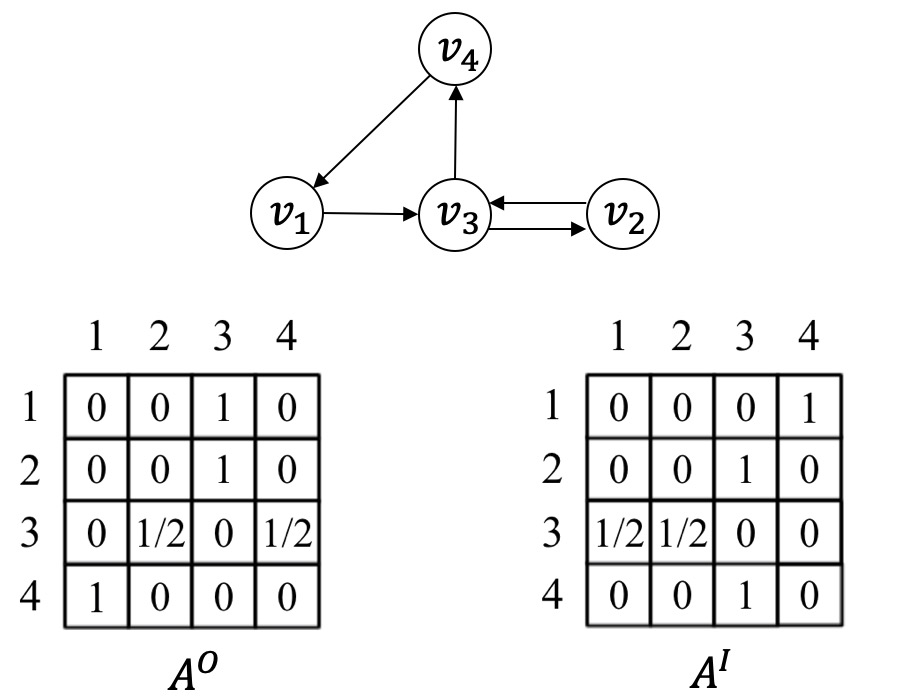}
    \caption{A example of a session graph and the connection matrix $\mathbf{A}^{O}$ and $\mathbf{A}^{I}$}.
    \label{figao}
\end{figure}

\subsubsection{Node representation learning}
Based on $\mathcal{G}_{s}^{intra}$, the first step to obtaining node representations is to get the node's contextual information by aggregating information from other nodes to the target node. That aggregation can be defined as two parallel processes: in-degree nodes aggregation and out-degree nodes aggregation. Those two aggregation processes are dependent on two adjacency matrices $\mathbf{A}^{O} , \mathbf{A}^{I} \in \mathbb{R}^{n \times n}$, which denote weighted connections of out-degree and in-degree edges in the session graph respectively. For example, considering a session $\{v_1, v_3, v_2, v_3, v_4, v_1\}$, the corresponding intra-session graph $\mathcal{G}_{s}^{intra}$ and adjacency matrices are illustrated in Figure \ref{figao}. According to $\mathbf{A}^{O} , \mathbf{A}^{I}$, the process of aggregation for target node $v_i$ in graph $\mathcal{G}_{s}^{intra}$ can be denoted as follow:
\begin{equation} \label{eq2}
    \begin{split}
    \begin{aligned}
    \mathbf{a}^{O, t}_{i} &= \mathbf{A}_{i}^{O}\left(\left[\mathbf{v}_{1}^{t-1}, \ldots, \mathbf{v}_{n}^{t-1}\right] \mathbf{W}_{O}\right) + \mathbf{b}^{O}, \\
    \mathbf{a}^{I, t}_{i} &= \mathbf{A}_{i}^{I}\left(\left[\mathbf{v}_{1}^{t-1}, \ldots, \mathbf{v}_{n}^{t-1}\right] \mathbf{W}_{I}\right) + \mathbf{b}^{I}, \\
    \end{aligned}
    \end{split}
\end{equation}

\begin{figure*}   
  \begin{minipage}[t]{0.3\linewidth} 
    \centering   
    \includegraphics[width=2.5in]{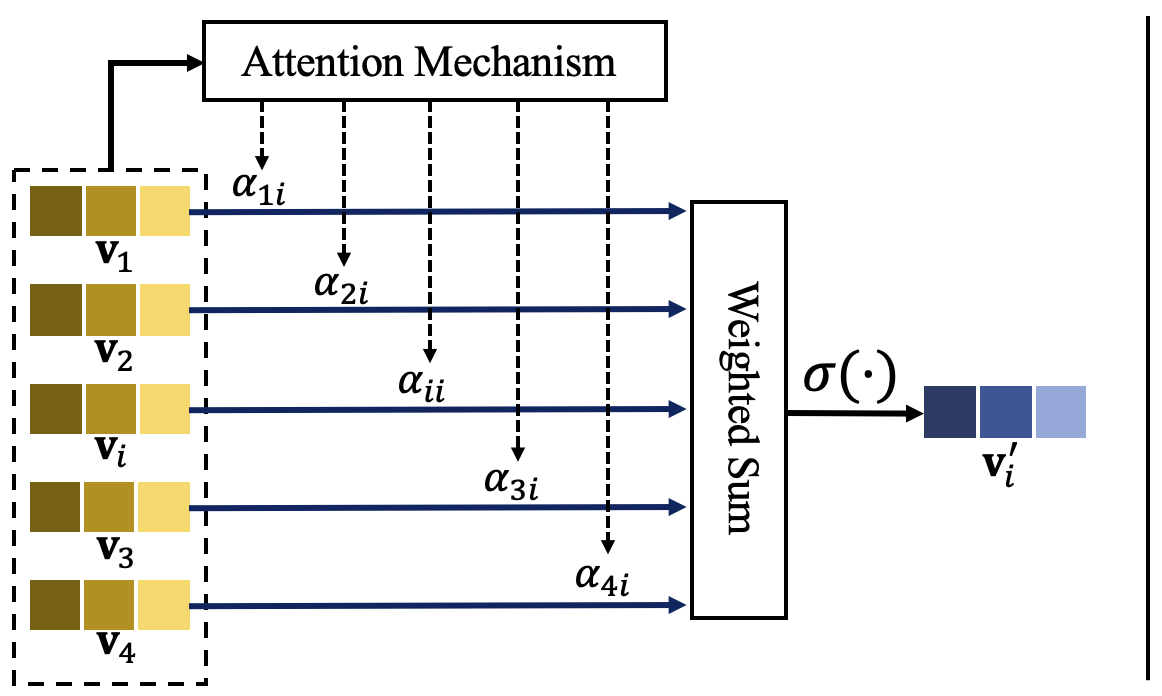}   
    \caption{Details of an attention head.}    
    \label{fig:side:a}   
  \end{minipage}%
  \hfill
  \begin{minipage}[t]{0.65\linewidth}   
    \centering   
    \includegraphics[width=4.5in]{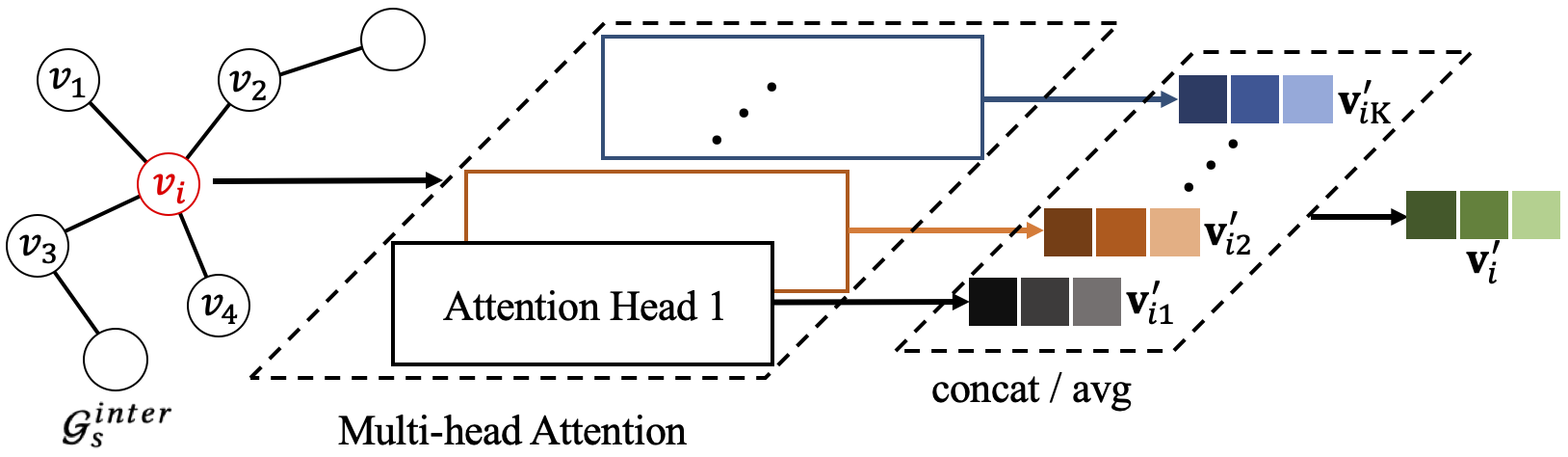}   
    \caption{The graphical model of the single convolutional layer using multi-head attention mechanism.} 
    \label{fig:side:b}   
  \end{minipage}   
\end{figure*} 

where $\mathbf{W}_{a}^{I}, \mathbf{W}_{a}^{O} \in \mathbb{R}^{d \times d}$ are parameter matrices. $\mathbf{b}^{I}, \mathbf{b}^{O} \in \mathbb{R}^{d}$ represent bias vectors. $\left[\mathbf{v}_{1}^{t-1}, \ldots, \mathbf{v}_{n}^{t-1}\right] \in \mathbb{R}^{n \times d} $ is the list of node vectors in the session $s$.  $\mathbf{A}_{i}^{O} , \mathbf{A}_{i}^{I} \in \mathbb{R}^{1 \times n}$ are two rows of elements in $\mathbf{A^{O}}$ and $\mathbf{A^{I}}$ respectively corresponding to the node $v_i$. After extracting the contextual information of out-degree nodes $\mathbf{a}^{O, t}_{i}$ and in-degree nodes $\mathbf{a}^{I, t}_{i}$, we combine them to get the final contextual information representation of node $v_i$, which can be denoted as $\mathbf{a}^{t}_{i}$, by the following operation:

\begin{equation} \label{eq3}
    \begin{split}
    \mathbf{a}^{t}_{i} = \left[\mathbf{a}^{O, t}_{i} || \mathbf{a}^{I, t}_{i}\right],
    \end{split}
\end{equation}
where $||$ represents the concatenation operation. 

After obtaining $\mathbf{a}^{t}_{i}$, the second step is to feed $\mathbf{a}^{t}_{i}$ and the previous hidden state $\mathbf{v}_i^{t-1}$ into gated update functions to update the hidden state of node $i$, where the update functions are demonstrated as follows:

\begin{equation} \label{eq4}
    \begin{split}
    \begin{aligned} 
    \mathbf{z}_{i}^{t} &=\sigma\left(\mathbf{W}_{z} \mathbf{a}_{i}^{t}+\mathbf{U}_{z} \mathbf{v}_{i}^{t-1}\right) \\ 
    \mathbf{r}_{i}^{t} &=\sigma\left(\mathbf{W}_{r} \mathbf{a}_{i}^{t}+\mathbf{U}_{r} \mathbf{v}_{i}^{t-1}\right) \\ 
    \widetilde{\mathbf{v}}_{i}^{t} &=\tanh \left(\mathbf{W}_{o} \mathbf{a}_{i}^{t}+\mathbf{U}_{o}\left(\mathbf{r}_{i}^{t} \odot \mathbf{v}_{i}^{t-1}\right)\right) \\ 
    \mathbf{v}_{i}^{t} &=\left(1-\mathbf{z}_{i}^{t}\right) \odot \mathbf{v}_{i}^{t-1}+\mathbf{z}_{i}^{t} \odot \widetilde{\mathbf{v}}_{i}^{t} 
    \end{aligned}
    \end{split}
\end{equation}

where $\mathbf{z}_{i}^{t}$ and $\mathbf{r}_{i}^{t}$ are the reset and update gates respectively, which decide what information to be preserved and discarded respectively. $\sigma(\cdot)$ represents the logistic sigmoid function and $\odot$ denotes element-wise multiplication. The whole process is like a typical GRU-based updating that integrating information from other nodes and previous states to update the current hidden state of the target node. When the update process for all nodes in the graph is finished, we obtain the final vector representation of each node.

\subsubsection{Intra-session representation generating}
After feeding session graph $\mathcal{G}_{s}^{intra}$ into the gated graph neural networks, we obtain the updated vectors of all nodes in session $s$. To alleviate the random interests drifts caused by users' unintended clicks, we combine both users' long-term preference and current interests of the session to generate the final intra-session representation. For the session $s = \{v_1, v_2, ... , v_n\}$, we use the last click item's embedding to represent user's current interests, i.e. $\mathbf{s}_\mathrm{l} = \mathbf{v}_n$. Then according to current interests $\mathbf{s}_\mathrm{l}$, we aggregate all node vectors in $s$ to obtain the long-term preference representation $\mathbf{s}_\mathrm{g}$ by adopting a soft-attention mechanism. Specifically, we derive $\mathbf{s}_\mathrm{g}$ by the following calculation:

\begin{equation} \label{eq5}
    \begin{split}
    \begin{array}{l}{\alpha_{i}=\mathbf{q}^{T} \sigma\left(\mathbf{W}_{1} \mathbf{v}_{n}+\mathbf{W}_{2} \mathbf{v}_{i}+\mathbf{b}\right)} \\
    {\mathbf{s}_{\mathrm{g}}=\sum_{i=1}^{n} \alpha_{i} \mathbf{v}_{i}}\end{array}
    \end{split}
\end{equation}
where $\mathbf{q}^{T} \in \mathbb{R}^{d}$ and $\mathbf{W}_{1}, \mathbf{W}_{2} \in \mathbb{R}^{d \times d}$ are learnable weighted parameters. 

Finally, we combine $\mathbf{s}_\mathrm{g}$ and $\mathbf{s}_\mathrm{l}$ to generate $\mathbf{s}^{intra}$. Technically, we first concatenate two interests $\mathbf{s}_\mathrm{g}$ and $\mathbf{s}_\mathrm{l}$, then use a linear transformation to compress the concatenation:
\begin{equation} \label{eq6}
    \begin{split}
    \mathbf{s}^{intra}=\mathbf{W}_{3}\left[\mathbf{s}_\mathrm{l} || \mathbf{s}_\mathrm{g}\right]
    \end{split}
\end{equation}
where $\mathbf{W}_{3} \in \mathbb{R}^{d \times 2 d}$ transfers the concatenation vectors latent space from $\mathbb{R}^{2d}$ to $\mathbb{R}^{d}$.

$\mathbf{s}^{intra}$ means the final representation of the current session, which only contains information within a single session, namely intra-session representation.

\subsection{Inter-session Module}

The aim of Inter-session Module is to integrate inter-session item-level interactions into the collaborative information for the current session based on the inter-session graph $\mathcal{G}^{inter}_s$. Same as Intra-session Module, the process of Inter-session Module can also be divided into the following two: \emph{Node representation learning} and \emph{Intra-session representation generating}, which are demonstrated as below.

\subsubsection{Node representation learning}
Because the graph is relatively large and the importance of nodes is unevenly distributed (not all items in neighbor sessions are equally related to the current session), assigning each node with a fixed normalized weight is not the best choice when aggregating neighbors nodes. Therefore, we introduce attention mechanisms \citep{Velickovic17} when modeling the complex interactions between $s$ and $N_s$. The attention mechanism can adaptively assign different weights to different neighbor nodes thus decreasing noises caused by less relevant items.

For the set of all item node vectors $V_s^{inter} = [\mathbf{v}_1, \mathbf{v}_2, ..., \mathbf{v}_N]$, where $N$ denotes the number of nodes in $\mathcal{G}_{s}^{inter}$, the shared self-attention mechanism $a$ is applied to every node to compute attention coefficients:

\begin{equation} \label{eq7}
    \begin{split}
    e_{i j}=a\left(\mathbf{W} \mathbf{v}_{i}, \mathbf{W} \mathbf{v}_j\right) = \operatorname{LeakyReLU}\left(\mathbf{a}^{T}\left[\mathbf{W} \mathbf{v}_{i} \| \mathbf{W} \mathbf{v}_{j}\right]\right)
    \end{split}
\end{equation}
where $\mathbf{W} \in \mathbb{R}^{d \times d}$ represents a shared weight matrix applied to every node vector and $e_{i j}$ indicates the importance of the node $j$’s vector to the node $i$. The attention mechanism $a$ applies a single-layer feedforward neural network with a weight vector $\mathbf{a} \in \mathbb{R}^{2 d}$. In addition, the LeakyReLU with the negative input slope $\alpha = 0.2$ is applied in the attention mechanism. In our experiment, we consider the first-order neighbors of $i$ (including $i$) in a single layer. So we only compute $e_{ij}$ for nodes $j \in \mathcal{N}_{i}$, where $\mathcal{N}_{i}$ denotes some neighborhoods of node $i$ in the graph (including $i$).

Then we normalize attention coefficients using the softmax function:

\begin{equation} \label{eq8}
    \begin{split}
    \alpha_{i j}=\operatorname{softmax}_{j}\left(e_{i j}\right)=\frac{\exp \left(e_{i j}\right)}{\sum_{k \in \mathcal{N}_{i}} \exp \left(e_{i k}\right)}
    \end{split}
\end{equation}



To stabilize the learning process of self-attention, we apply multi-head attention \citep{Velickovic17} in Inter-session Module. Specifically, we use $K$ independent attention heads to extract $K$ different latent vectors of node $i$. Details of each attention head are illustrated in Figure \ref{fig:side:a}. Then the module concatenates those node vectors as an output. So the update process of node $i$ is defined by the following operation:
\begin{equation} \label{eq10}
    \begin{split}
    \mathbf{v}_{i}^{\prime}=\|_{k=1}^{K} \sigma\left(\sum_{j \in \mathcal{N}_{i}} \alpha_{i j}^{k} \mathbf{W}^{k} \mathbf{v}_{j}\right)
    \end{split}
\end{equation}

where $\alpha_{i j}^{k}$ represents normalized attention coefficients computed by the $k$-th attention mechanism $a^k$, $\|_{k=1}^{K}$ represents the concatenation for all attention mechanism outputs, and $\sigma(\cdot)$ represents Sigmoid function. $\mathbf{v}_{i}^{\prime} \in \mathbb{R}^{K \times d}$ denotes the updated vector of node $i$.

Whereas for the output layer, because we need to reduce the dimension of the node vector form $\mathbb{R}^{K \times d}$ to $\mathbb{R}^{d}$, we use averaging instead of combining. So Equation \ref{eq10} can be rewritten as:

\begin{equation} \label{eq11}
    \begin{split}
    \mathbf{v}_{i}^{\prime}=\sigma\left(\frac{1}{K} \sum_{k=1}^{K} \sum_{j \in \mathcal{N}_{i}} \alpha_{i j}^{k} \mathbf{W}^{k} \mathbf{v}_{j}\right)
    \end{split}
\end{equation}

The structure of a single convolutional layer with multi-head attention is demonstrated in Figure \ref{fig:side:b}.

According to the experiment result, the number of convolutional layers is set to two and the number of attention heads $K$ is set to eight for both layers to obtain the best result.

After updating, each node in session $s$ aggregates the information of its neighbor nodes in multiple sessions. Thus the collaborative information within related items among different sessions is encoded into each node's representation. In order to distinguish the vector representation obtained by Intra-session Module and Inter-session Module, we use $\overline{\mathbf{v}}_i$ to denote the updated latent vector of node $i$.

\begin{table*}
\centering
\scriptsize
\caption{Statistics of datasets used in our experiments.}\label{tab1}
\resizebox{0.9\textwidth}{!}{
\begin{tabular}{l ccccc}
    \toprule
    \textbf{Datasets} & \# of clicks &  \# of training sessions &  \# of testing sessions  &  \# of items & average length \\
    \midrule

    \textbf{YOOCHOOSE 1/64}  
             &   565,552   &   375,043  &   55,405   &   17,319   &   6.07           \\
    \textbf{YOOCHOOSE 1/4}     
             &   7,980,529   &   5,969,416  &   55,872   &   30,638   & 5.71            \\
    \textbf{Diginetica}    
             &   982,961   &   719,470  &   68,977   &   43,097   &   5.12           \\
    
    \bottomrule
\end{tabular}}
\end{table*}

\subsubsection{Inter-session representation generating}
The process of generating inter-session representation is similar to that in Intra-session Module. For session $s = \{v_1, v_2, ... , v_n\}$, we use the last click item's updated node vectors $\overline{\mathbf{v}}_n$ to represent the user's current interest $\overline{\mathbf{s}}_\mathrm{l}$ with collaborative information, thus  $\overline{\mathbf{s}}_\mathrm{l}=\overline{\mathbf{v}}_n$. Then the same soft-attention mechanism is adopted to aggregate all updated node vectors in $s$ to obtain long-term preference $\overline{\mathbf{s}}_\mathrm{g}$ with collaborative information, the calculation formation is similar as Equation \ref{eq5}, where $\mathbf{s}_\mathrm{g} =\overline{\mathbf{s}}_\mathrm{g}$, $\mathbf{v}_\mathrm{n} =\overline{\mathbf{v}}_n$.

Finally, a linear transformation is applied on the concatenation of two types of neighborhood information $\overline{\mathbf{s}}_\mathrm{l}$ and $\overline{\mathbf{s}}_\mathrm{g}$ to compute the inter-session information representation:
\begin{equation} \label{eq12}
    \begin{split}
    \mathbf{s}^{inter}=\overline{\mathbf{W}}_{3}\left[\overline{\mathbf{s}}_\mathrm{l} || \overline{\mathbf{s}}_\mathrm{g}\right]
    \end{split}
\end{equation}

$\mathbf{s}^{inter}$ represents the inter-session representation generated from the current session's neighbors that should have high relevance to the current session.

\subsection{Intra-and Inter-Session Representations Combination}
After obtaining the intra-session representation $\mathbf{s}^{intra}$ and inter-session representation $\mathbf{s}^{inter}$, the last step for the final session representation is to combine them. Inspired by \citep{wang2019CSRM}, we use fusion gating mechanism to obtain the final session representation $\mathbf{s}^{h}$:

\begin{equation} \label{eq13}
    \begin{split}
    \begin{aligned}
    \boldsymbol{f} &= \sigma\left(\boldsymbol{W}_{f}^1 \mathbf{s}^{inter}+\boldsymbol{W}_{f}^2 \mathbf{s}^{intra} + \boldsymbol{b_f}\right) \\
    \mathbf{s}^h &= \boldsymbol{f} \mathbf{s}^{inter}+\left(1-\boldsymbol{f}\right) \mathbf{s}^{intra}
    \end{aligned}
    \end{split}
\end{equation}

where $\boldsymbol{W}_{f}^1$, $\boldsymbol{W}_{f}^2$ denote the weight matrices in fusion gate and $\boldsymbol{b}_{f}$ represents the bias vector. And $\mathbf{s}^{h}$ is the final session representation.

\subsection{Prediction Module and Objective Function}
After obtaining the final session representation $\mathbf{s}^{h}$, we use it to multiply each candidate item vector $\mathbf{v}_{\mathrm{i}} \in \mathbf{V}$ to generate recommendation score $\hat{\mathbf{z}_{i}}$ for corresponding item:
\begin{equation} \label{eq14}
    \begin{split}
    \hat{\mathbf{z}}_{i}=\mathbf{v}_{i}^{T} \mathbf{s}^{h}
    \end{split}
\end{equation}

Then we apply a softmax function to generate the output vector of the model $\hat{\mathbf{y}}$:
\begin{equation} \label{eq15}
    \begin{split}
    \hat{\mathbf{y}}=\operatorname{softmax}(\hat{\mathbf{z}})
    \end{split}
\end{equation}
where $\hat{\mathbf{z}} \in \mathbb{R}^{|I|}$ represents the recommendation scores over all candidate items and $\hat{\mathbf{y}} \in \mathbb{R}^{|I|}$ denotes the probabilities of items becoming the next-click item in session $s$.

In the training process, we apply Cross-entropy as the loss function:
\begin{equation} \label{eq16}
    \begin{split}
    \mathcal{L}(\hat{\mathbf{y}})=-\sum_{i=1}^{|I|} \mathbf{y}_{i} \log \left(\hat{\mathbf{y}}_{i}\right)+\left(1-\mathbf{y}_{i}\right) \log \left(1-\hat{\mathbf{y}}_{i}\right)
    \end{split}
\end{equation}
where $\mathbf{y}$ denotes the one-hot encoding vector of the ground truth item.

The proposed I3GN is trained by Back-Propagation Through Time (BPTT) algorithm in the learning process.

\section{Experiments}
In this section, we describe the information of datasets used in experiments and introduce the baseline methods used for comparison.
\subsection{Datasets}

To evaluate the efficiency of our proposed method, we conduct experiments on two real-world datasets: YOOCHOOSE\footnote{http://2015.recsyschallenge.com/challenge.html} dataset and Diginetica\footnote{http://cikm2016.cs.iupui.edu/cikm-cup} dataset. The YOOCHOOSE dataset is released by the RecSys challenge 2015, which records click sequences (item views, purchases) for a period of six months. The Diginetica dataset is published by CIKM Cup 2016, in which we only select the transaction data for experiments. 

We filter out sessions of length one and items that appear less than five times for both datasets as same as previous studies \citep{liu2018stamp, wu2019srgnn}. Furthermore, we use the last one day in YOOCHOOSE and last seven days in Diginetica to generate the test data. Because collaborative filtering idea-based methods cannot recommend an item which has not appeared before \citep{hidasi2015session}, we filter out items from test set which do not appear in the training set. According to previous studies \citep{liu2018stamp, wu2019srgnn}, we use the most recent 1/4 and 1/64 of training sessions in YOOCHOOSE, which make up the YOOCHOOSE 1/64 and YOOCHOOSE 1/4 datasets respectively. Similar to \citep{wu2019srgnn}, data augmentation is applied to preprocess the data. Specifically, we augment the data by splitting input sessions. For example, for an input session $s=[v_1, v_2, ... ,v_n]$, we generate the sub-sessions and their corresponding labels $([v_1], v_2), ([v_1, v_2], v_3)$, ..., $([v_1, v_2, ...,v_{n-1}], v_n)$. We also sort all sessions in chronological order for all datasets.

The statistics of datasets are shown in Table \ref{tab1}.

\subsection{Baseline}
We compare proposed method with following representative and state-of-the-art methods as baselines to evaluate the performance:
\begin{itemize}
\item {\textbf{POP}: A model that always recommends the most popular items in the training set.}
\item {\textbf{Item-KNN} \citep{sarwar2001item}: A traditional model that recommends items based on the similarity between the existing items in the session.}
\item {\textbf{FPMC} \citep{rendle2010factorizing}: A hybrid model that combines Matrix Factorization and Markov Chain for next-basket recommendation.}
\item {\textbf{BPR-MF} \citep{rendle2009bpr}: A widely used matrix factorization method, which optimizes a pairwise ranking objective function with Bayesian Personalized Ranking loss.}
\item {\textbf{SKNN} \citep{bonnin2014sknn}: A neighborhood-based method considering the integrity of sessions.}    
\item {\textbf{GRU4REC} \citep{hidasi2015session}: An RNN-based SRS model. It employs GRU units and the session-parallel mini-batch training process.}
\item {\textbf{NARM} \citep{li2017neural}: This model employs RNNs with the attention mechanism to capture the user's main purpose and sequential behavior and combines them to make recommendations.}
\item {\textbf{STAMP} \citep{liu2018stamp}: This model uses the last click to represent the short-term interest and utilize the attention mechanism to capture the user's long-term interest. Then it combines them to make recommendations.}
\item {\textbf{SR-GNN} \citep{wu2019srgnn}: A model uses Graph Neural Networks to generate latent vectors of items and make recommendations with attention mechanisms.}
\item {\textbf{KNN-RNN} \citep{jannach2017recurrent}: A hybrid model that weightedly combine GRU4REC with SKNN to get a better result.}
\item {\textbf{CSRM} \citep{wang2019CSRM}: A hybrid neural network-based framework which takes session-level collaborative information into account.}
\end{itemize}

\begin{table*}
\centering
\scriptsize
\caption{Performance comparison of I3GN with baseline methods.}\label{tab2}
\resizebox{0.9\textwidth}{!}{
\begin{tabular}{ccccccccccccc}
        \toprule
    \multirow{4}{*}{\textbf{Methods}} & \multicolumn{4}{c}{\textbf{YOOCHOOSE 1/64}} & \multicolumn{4}{c}{\textbf{YOOCHOOSE 1/4}} & \multicolumn{4}{c}{\textbf{DIGINETICA}}\\
    \cmidrule(r){2-5} \cmidrule(r){6-9} \cmidrule(r){10-13} 
    
     & \multicolumn{2}{c}{\textbf{MRR}} & \multicolumn{2}{c}{\textbf{Recall}} & \multicolumn{2}{c}{\textbf{MRR}} & \multicolumn{2}{c}{\textbf{Recall}} & \multicolumn{2}{c}{\textbf{MRR}} & \multicolumn{2}{c}{\textbf{Recall}}\\
    \cmidrule(r){2-3} \cmidrule(r){4-5}  \cmidrule(r){6-7}  \cmidrule(r){8-9}  \cmidrule(r){10-11} \cmidrule(r){12-13} 
      & @5 & @10 & @5 & @10 & @5 & @10 & @5 & @10 & @5 & @10 & @5 & @10 \\
     \midrule
    \textbf{POP}      
             & 1.36 & 1.51      & 3.29 & 4.59  
             & 0.20 & 0.27      & 0.37 & 0.85 
             & 0.19 & 0.22      & 0.39 & 0.63  \\
    \textbf{Item-KNN}    
             & 19.97 & 21.38      & 32.80 & 43.39  
             & 19.57 & 21.08      & 32.07 & 43.38 
             & 8.05 & 8.95      & 14.47 & 21.30  \\
    \textbf{SKNN} 
             & 22.55 & 24.31      & 39.22 & 52.34  
             & 22.52 & 24.29      & 39.17 & 52.38 
             & 16.37 & 17.79      & 27.46 & 38.12  \\
    \textbf{FPMC}     
             & 19.76 & 20.85      & 29.61 & 37.81 
             & 16.69 & 17.90      & 26.79 & 35.96 
             & 15.84 & 16.09      & 18.55 & 20.45  \\        
    \textbf{BPR-MF}   
             & 17.77 & 18.35      & 24.86 & 29.16  
             & 15.89 & 16.12      & 20.15 & 21.85 
             & 13.39 & 13.50      & 16.69 & 17.53  \\
    \midrule         
    \textbf{GRU4REC}  
             & 24.60 & 26.48      & 39.22 & 52.34  
             & 20.11 & 21.78      & 34.55 & 47.12 
             & 6.69 & 7.69      & 12.98 & 20.52  \\
    \textbf{NARM} 
             & 26.21 & 27.97      & 44.34 & 57.50  
             & 26.08 & 28.10      & 44.34 & 57.83 
             & 25.02 & 26.53      & 40.67 & 51.91  \\
    \textbf{STAMP}    
             & 27.26 & 28.92      & 45.69 & 58.07 
             & 27.47 & 29.24      & 46.39 & 59.62 
             & 25.21 & 26.69      & 41.04 & 52.07  \\
    \textbf{SR-GNN}   
             & 28.01 & 29.97      & 46.49 & 60.33  
             & 29.34 & 31.08      & 48.15 & 61.06 
             & 25.56 & 26.82      & 41.11 & 51.47  \\
    \midrule
    \textbf{KNN-RNN}  
             & 25.39 & 27.26      & 43.15 & 57.03  
             & 22.00 & 23.60      & 37.17 & 49.06 
             & 10.42 & 11.51      & 13.40 & 21.06  \\
    \textbf{CSRM}     
             & 27.84 & 29.62      & 46.76 & 60.06 
             & 28.91 & 29.12      & 47.98 & 60.28 
             & 25.17 & 26.64      & 41.36 & \textbf{52.89}  \\
    \textbf{I3GN}    
             & \textbf{28.67} & \textbf{30.44}      & \textbf{47.32} & \textbf{60.41}  
             & \textbf{29.53} & \textbf{31.28}      & \textbf{48.33} & \textbf{61.29} 
             & \textbf{26.30} & \textbf{27.84}      & \textbf{41.81} & 52.66  \\
    
    \bottomrule
\end{tabular}}
\end{table*}

\subsection{Evaluation Metrics}
To evaluate the performance of the proposed method, we adopt the following two common metrics in our experiments:
\begin{itemize}
\item {\textbf{Recall@$N$}: Recall@$N$ is a common metric to evaluate the performance of SRS model. Recall@$N$ is the proportion of cases having the desired item amongst the top-$N$ items in all test cases.}
\item {\textbf{MRR@$N$}: MRR@$N$(Mean Reciprocal Rank) is the average of reciprocal ranks of the desired items. The reciprocal rank is set to zero if the value of rank exceeds $N$. MRR is especially important to measure the performance of SRS because it considers the order of recommendation results and users tend to focus on higher-ranked items.}
\end{itemize}

Because most users are only interested in viewing recommendations on the first page of real application scenarios (e.g., web sites of e-commerce), the relevant item should be amongst the first few items in the recommendation list \citep{hu2017diversifying, quadrana2017personalizing}. So we report values of all metrics at $N$=$\{5, 10\}$. 

\begin{table*}
\centering
\scriptsize
\caption{Performance comparison of I3GN with different level session representation on three datasets.}\label{tab3}
\resizebox{0.9\textwidth}{!}{
\begin{tabular}{ccccccccccccc}
        \toprule
    \multirow{4}{*}{\textbf{Methods}} & \multicolumn{4}{c}{\textbf{YOOCHOOSE 1/64}} & \multicolumn{4}{c}{\textbf{YOOCHOOSE 1/4}} & \multicolumn{4}{c}{\textbf{DIGINETICA}}\\
    \cmidrule(r){2-5} \cmidrule(r){6-9} \cmidrule(r){10-13} 
    
     & \multicolumn{2}{c}{\textbf{MRR}} & \multicolumn{2}{c}{\textbf{Recall}} & \multicolumn{2}{c}{\textbf{MRR}} & \multicolumn{2}{c}{\textbf{Recall}} & \multicolumn{2}{c}{\textbf{MRR}} & \multicolumn{2}{c}{\textbf{Recall}}\\
    \cmidrule(r){2-3} \cmidrule(r){4-5}  \cmidrule(r){6-7}  \cmidrule(r){8-9}  \cmidrule(r){10-11} \cmidrule(r){12-13} 
      & @5 & @10 & @5 & @10 & @5 & @10 & @5 & @10 & @5 & @10 & @5 & @10 \\
     \midrule

   $ \textbf{I3GN}_{\mathrm{inter}}$  
             & 26.87 & 28.79      & 44.86 & 57.76  
             & 26.99 & 28.69      & 45.10 & 57.65 
             & 25.28 & 26.67      & 40.22 & 50.62  \\
    $\textbf{I3GN}_{\mathrm{intra}}$ 
             & 28.01 & 29.97      & 46.49 & 60.33  
             & 29.34 & 31.08      & 48.15 & 61.06 
             & 25.56 & 26.82      & 41.11 & 51.47  \\
   \textbf{I3GN}    
             & \textbf{28.67} & \textbf{30.44}      & \textbf{47.32} & \textbf{60.41}  
             & \textbf{29.53} & \textbf{31.28}      & \textbf{48.33} & \textbf{61.29} 
             & \textbf{26.30} & \textbf{27.84}      & \textbf{41.81} & \textbf{52.66}  \\
    
    \bottomrule
\end{tabular}
}
\end{table*}

\subsection{Parameter Setup}
In our experiments, we set the embedding dimension of items as 100 on two YOOCHOOSE datasets and 50 on Diginetica dataset. We use a Gaussian distribution with a mean of zero and a standard deviation of 0.1 to initialize model parameters. We also adopt the mini-batch Adam optimizer to optimize parameters, where the initial learning rate is set to 0.001. The batch size is set to 128 on both YOOCHOOSE 1/64 and Diginetica, and it is 256 on YOOCHOOSE 1/4. For parameters of Inter-session Module, we decay the learning rate by 0.1 every five epochs, while the other modules in I3GN decay the learning rate by 0.1 every three epochs. We set the number of nearest neighbors $k = 120$ in the neighbor sessions retrieval according to the experiment results. For a fair comparison, on each dataset, we unify the dimension of embedding for all baselines and set the number of neighbors in CSRM to the same as ours. We use PyTorch to implement our model where graph models are carried out by PyTorch Geometric library \citep{fey2019fast}. The model is trained on a Geforce Titan V GPU.

\section{Results and analyses}
\begin{table*}
\centering
\scriptsize
\caption{Performance comparison of I3GN with different graph modeling strategies.}\label{tab4}
\resizebox{0.9\textwidth}{!}{
\begin{tabular}{ccccccccccccc}
        \toprule
    \multirow{4}{*}{\textbf{Methods}} & \multicolumn{4}{c}{\textbf{YOOCHOOSE 1/64}} & \multicolumn{4}{c}{\textbf{YOOCHOOSE 1/4}} & \multicolumn{4}{c}{\textbf{DIGINETICA}}\\
    \cmidrule(r){2-5} \cmidrule(r){6-9} \cmidrule(r){10-13} 
    
     & \multicolumn{2}{c}{\textbf{MRR}} & \multicolumn{2}{c}{\textbf{Recall}} & \multicolumn{2}{c}{\textbf{MRR}} & \multicolumn{2}{c}{\textbf{Recall}} & \multicolumn{2}{c}{\textbf{MRR}} & \multicolumn{2}{c}{\textbf{Recall}}\\
    \cmidrule(r){2-3} \cmidrule(r){4-5}  \cmidrule(r){6-7}  \cmidrule(r){8-9}  \cmidrule(r){10-11} \cmidrule(r){12-13} 
      & @5 & @10 & @5 & @10 & @5 & @10 & @5 & @10 & @5 & @10 & @5 & @10 \\
     \midrule
    \textbf{I3GN}-$avg$  
             & 24.69 & 26.84      & 41.72 & 55.12  
             & 25.38 & 27.23      & 43.12 & 56.91 
             & 20.28 & 21.87      & 34.80 & 46.65  \\
    \textbf{I3GN}-$\alpha \& \beta$  
             & 27.76 & 29.48      & 46.02 & 58.89  
             & 28.50 & 30.25      & 47.10 & 60.12 
             & 24.86 & 26.28      & 39.93 & 50.55  \\
    \textbf{I3GN}-$\beta$  
             & 28.04 & 29.77      & 46.62 & 59.53
             & 28.74 & 30.48      & 47.78 & 60.70 
             & 25.45 & 26.87      & 40.48 & 51.12  \\
    \textbf{I3GN}-$\alpha$  
             & 28.49 & 30.10      & 47.05 & 60.06  
             & 29.18 & 30.92      & 47.95 & 60.85 
             & 26.12 & 27.45      & 41.53 & 52.24  \\
    \textbf{I3GN}    
             & \textbf{28.67} & \textbf{30.44}      & \textbf{47.32} & \textbf{60.41}  
             & \textbf{29.53} & \textbf{31.28}      & \textbf{48.33} & \textbf{61.29} 
             & \textbf{26.30} & \textbf{27.84}      & \textbf{41.81} & \textbf{52.66}  \\
    
    \bottomrule
\end{tabular}}
\end{table*} 
In this section we compare the proposed model with other state-of-the-art methods, then we conduct detailed analyses of our model under different experimental settings.

\subsection{Comparison with baseline methods}
The experimental results of all methods in top-5 and top-10 recommendation on YOOCHOOSE and Diginetica datasets are illustrated in Table \ref{tab2}, and the following observations stand out:
\begin{itemize}

    \item {Two KNN-based methods: Item-KNN and SKNN considerably outperform other conventional baseline methods. This proves the effectiveness of adopting inter-session collaborative information on recommendations. Furthermore, SKNN takes the entire current session into consideration when calculating similarity while Item-KNN only considers the last item in the current session and ignores session contextual information. So SKNN achieves a better result.}
    \item {All of the neural network-based methods distinctly outperforms other conventional recommendation methods, demonstrating the superiority of adopting deep learning technology to make recommendations. The key reason for this may be RNN's ability to process sequentiality and thus model the intra-session item-level interactions.}
    \item {By comparing the performance of the original model and its neighborhood-enhanced version (e.g., GRU4REC and KNN-RNN, NARM and CSRM), we can observe that utilizing neighborhood information can enhance model performance. This result confirms the effectiveness of combining different scopes in session-based recommendations.}
    \item {On the whole, graph-based methods (SR-GNN and I3GN) outperform RNN-based methods (GRU4REC, NARM, STAMP, KNN-RNN, and CSRM). This indicates that it is important for SRS to explicitly model contextual item-level interactions because GNN can easily extract those by aggregating information among multiple nodes, but RNN can only deal with unidirectional transitions between consecutive items.}
    \item {Finally, our proposed I3GN obtains the best performance in almost every experiment, which validates that taking inter-session item-level interactions into account is beneficial. Although the performance of I3GN on Diginetica is lower than CSRM in terms of Recall@10, our model performs better than CSRM under stricter rules (e.g., the evaluation metric of Recall@5 on all datasets).
    }

\end{itemize}

\subsection{Influence of Inter-session Module}
We further analyze the effort of utilizing inter-session information. Two downgraded versions of our model are proposed to compare with I3GN: $\mathrm{I3GN}_\mathrm{inter}$ is a version of I3GN without Intra-session Module, and $\mathrm{I3GN}_\mathrm{intra}$ refers to I3GN without Inter-session Module. $\mathrm{I3GN}_\mathrm{intra}$ only models intra-session interactions and extracts information to make recommendations. Since the process of modeling the current session and obtaining the intra-session representation $\mathbf{s}^{intra}$ is identical to SR-GNN, we use the performance of SR-GNN to represent $\mathrm{I3GN}_\mathrm{intra}$.

$\mathrm{I3GN}_\mathrm{inter}$ ignores the intra-session information and extracts item-level interactions between the current session and its neighbor sessions directly. Table \ref{tab3} displays the experimental results on three datasets.

Results show that $\mathrm{I3GN}_\mathrm{inter}$ substantially outperforms $\mathrm{I3GN}_\mathrm{intra}$, which demonstrates that intra-session information plays a more crucial role when making recommendations. Moreover, the best performance obtained by I3GN reveals that combining two types is an effective strategy.

\subsection{Influence of the Number of Neighbors}
\begin{figure}[t]
    \centering
    \includegraphics[width=0.5\textwidth]{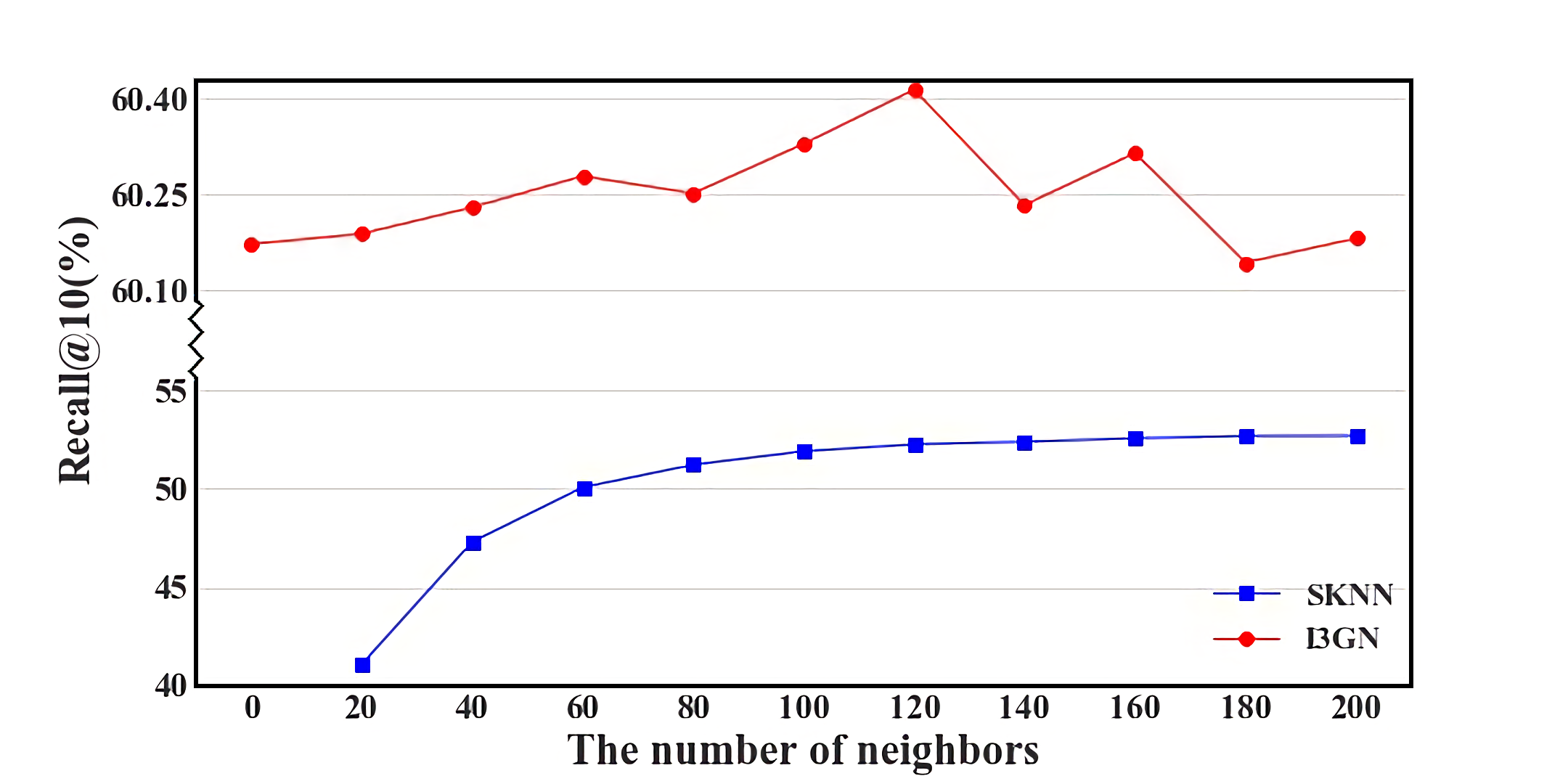}
    \caption{The Recall@10 of I3GN and SKNN with different number of neighbors on the YOOCHOOSE 1/64 dataset.}
    \label{fig-neigh}
\end{figure}

In this section, we vary the neighbor number $k$ to investigate its influence. We vary $k$ from zero to 200 and conduct our experiments on YOOCHOOSE 1/64 dataset. $k = 0$ means the neighbor sessions $N_s$ are eliminated and only the current session $s$ is used to build the graph in Inter-session Module, thus no information from other sessions could be influential. In order to better analyze the role of $k$, we adopt SKNN to compare it with I3GN. The parameters of SKNN are the same as the neighbors retrieval process in I3GN. Formally, given a session $s$, the recommendation score for each item $i$ is generated by the following computation: 
\begin{equation} \label{eq14}
    \begin{split}
    score_{\mathrm{SKNN}}(i, s)=\Sigma_{j \in N_{s}} \operatorname{sim}(s, j) \cdot 1_{\mathrm{j}}(i)
    \end{split}
\end{equation}
where the indicator function $1_\mathrm{j}(i)$ returns one if session $j$ contains item $i$ and 0 otherwise.
SKNN is unable to make recommendations without neighbor sessions, so we do not report the result of SKNN when $k = 0$. The results of Recall@10 with different $k$ are illustrated in Figure \ref{fig-neigh}. 

From Figure \ref{fig-neigh} we can observe that, with the increase of $k$, the performance of I3GN and SKNN are increased at first since the more neighbors there are for each session, the more information could be utilized to make recommendations. However, after $k = 120$, the performance of I3GN starts to drop and the improvement of SKNN is marginal. This result can be explained that when $k$ reaches a certain value, the benefits brought by additional growth of less similar neighbor sessions to SKNN gradually decrease, and extra noise sessions began to have negative effects on I3GN.

\subsection{Further Analysis of Inter-session Module}
To deeply understand the mechanism of Inter-session Module in I3GN, we further conduct experiments to analyze the efficacy of two pivotal components: graph modeling and attention mechanisms. Four variants of our model are proposed for comparisons: 
\begin{itemize}
    \item{I3GN-$avg$: This model uses an average pooling operation on all item node vectors $V_s^{inter}$ in neighborhood sessions to generate $\mathbf{s}^{inter}$, which is formulated as:
    \begin{equation} \label{eq1}
    \begin{split}
    \mathbf{s}^{inter}=\frac{1}{N} \sum_{i=1}^{N} \mathbf{v}_{i}
    \end{split}
    \end{equation}
where $N = |V_s^{inter}|$. I3GN-$avg$ represents the simplest way to employ collaborative information without graph structure modeling and attention mechanisms.
    }
    \item{I3GN-$\alpha$: This model uses a mean-based aggregation method instead of applying attention mechanisms while aggregating current session's neighbors. Each edge in $\mathcal{G}_s^{inter}$ has fixed normalized weight and the aggregation process of node $i$ can be formalized as:
    \begin{equation} \label{eq1}
    \begin{split}
    \mathbf{v}_{i}^{\prime}= \sigma\left(\sum_{j \in \mathcal{N}_{i}} \alpha_{i j} \mathbf{W}\mathbf{v}_{j}\right)
    \end{split}
    \end{equation}
    
    where $\alpha_{i j} = \frac{1}{|\mathcal{N}_i|}$.
    } 
    \item{I3GN-$\beta$: This model uses average pooling operation on current session's node vectors to replace soft attention mechanisms when generating long-term preference related neighbors information. So the calculation formation of $\overline{\mathbf{s}}_\mathrm{g}$ in Inter-session Module can be rewritten as:
    \begin{equation} \label{eq1}
    \begin{split}
    \overline{\mathbf{s}}_\mathrm{g}=\frac{1}{|s|} \sum_{i=1}^{|s|} \mathbf{v}_{i}
    \end{split}
    \end{equation}
    } 
    \item{I3GN-$\alpha  \& \beta$: This variant combines I3GN-$\alpha$ and I3GN-$\beta$, removing all attention mechanisms in Inter-session Module but reserving the graph structure.
    } 
\end{itemize}
The results of comparisons among I3GN and its variants are shown in Table \ref{tab4}. Observations of the results can be listed as follows:

\begin{itemize}
    \item {I3GN-$avg$ gets the worst performance in the experiments, the lack of graph structure and attention mechanisms leads to dramatic performance drops.}
    \item {Compare the performance of I3GN-$avg$ with I3GN-$\alpha \& \beta$, the adoption of the graph structure significantly increases the model performance. Besides, applications of attention mechanisms (I3GN-$\alpha$ and I3GN-$\beta$) further boost the performance. This result indicates that not all interacted items in the Inter-session Graph contribute equally to the current session, i.e., not all items from neighborhood sessions are relevant to users' interests.}
    \item {The performance of I3GN-$\alpha$ outperforms I3GN-$\beta$ in all experiments, we speculate the reason is that neighbor nodes of less important items in a current session $s$ may also be less important when making recommendations.}
    \item {I3GN makes full advantage of utilizing graph structure to extract complex item-level interactions between $s$ and its neighbors and applying attention mechanisms to alleviate the influence of less relevant items. Hence I3GN surpasses all variants of itself.}
\end{itemize}

\begin{figure}   
  \begin{minipage}[t]{0.5\linewidth} 
    \centering   
    \includegraphics[width=1.6in]{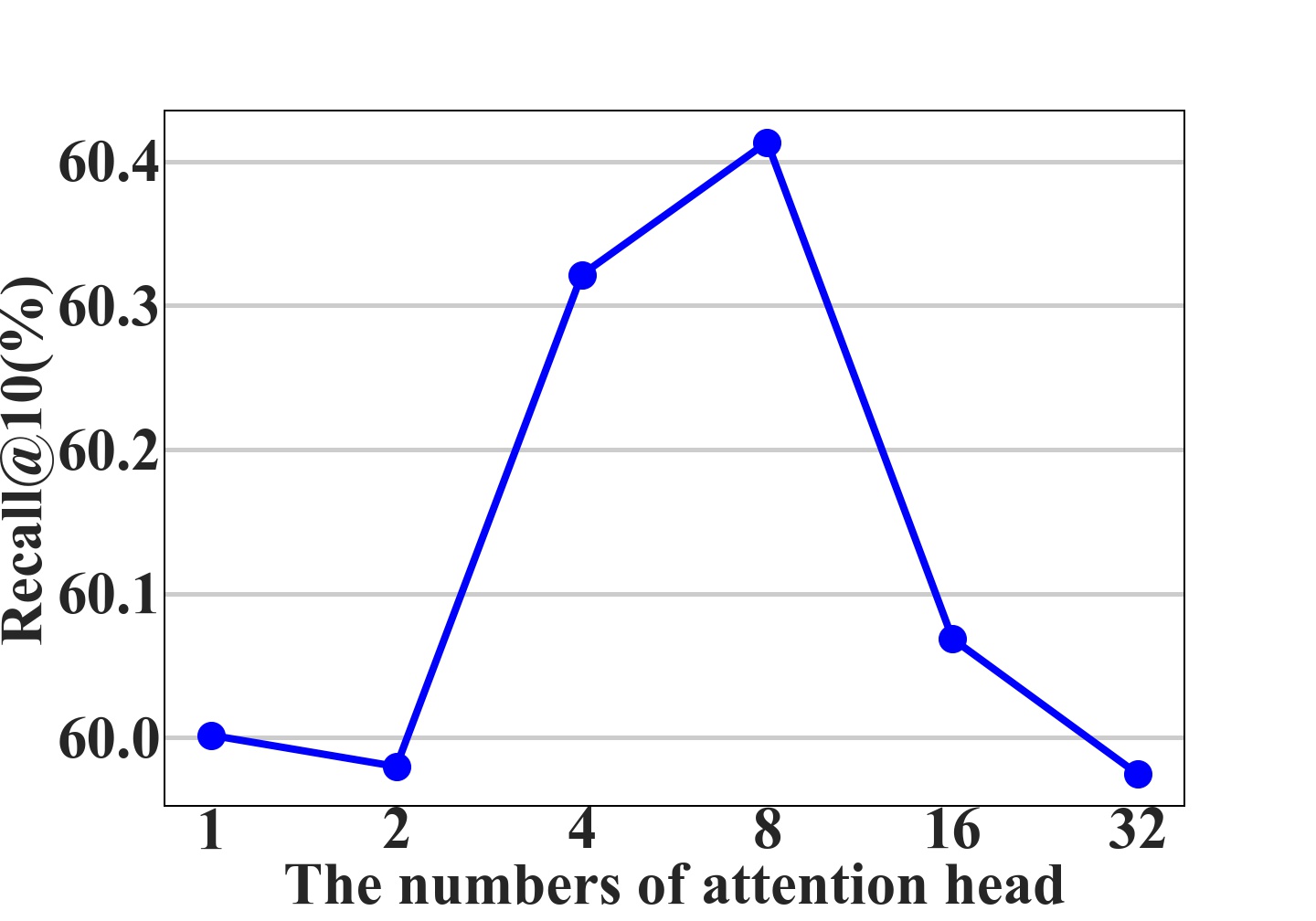}   
    \caption{The number of \\ attention heads.}   
    \label{figK}   
  \end{minipage}%
  \hfill
  \begin{minipage}[t]{0.5\linewidth}   
    \centering   
    \includegraphics[width=1.6in]{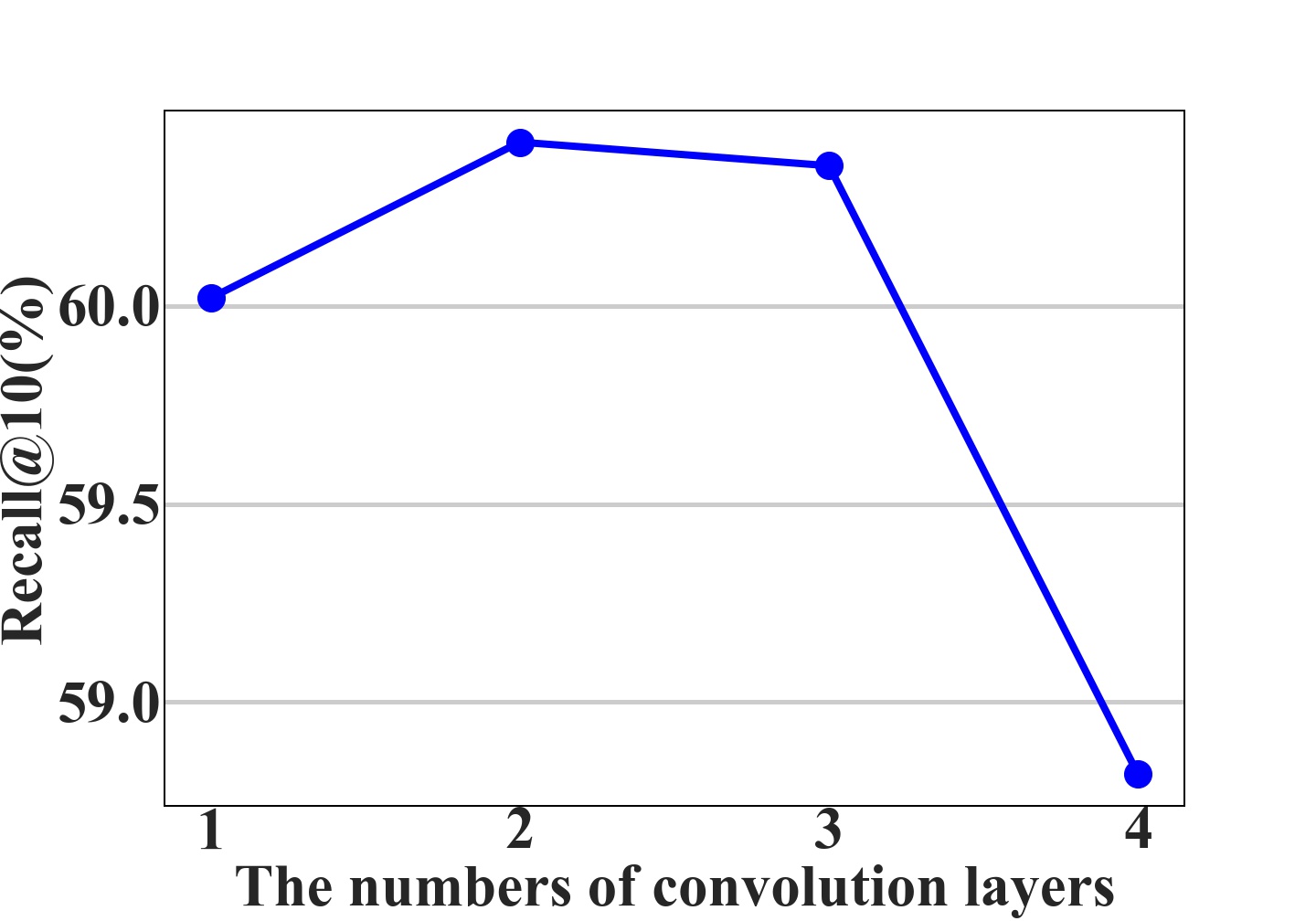}   
    \caption{The number of convolution layers.}   
    \label{figl}   
  \end{minipage}   
\end{figure} 

\subsection{Parameter sensitivity}
In order to study the effect of the number of attention heads $K$ and the number of graph convolutional layers in Inter-session Module, we vary $K$ from one to 32 and layer number from one to four, respectively. Other parameters are fixed in experiments. The results of Recall@10 on YOOCHOOSE 1/64 dataset are shown as Figure \ref{figK} and Figure \ref{figl}.  
From Figure \ref{figK} we can observe that as the number of attention heads increases, the performance of I3GN also increases because multi-head attention brings a more stable learning process. However, the performance gets worse after $K = 8$, this result may be due to possible overfitting. 

According to Figure \ref{figl}, the best performance of I3GN is achieved when the layer number of graph convolution is set to two. We assume that fewer layers could only encode limited information from neighborhood sessions and more layers may aggregate higher-order nodes which may be less relevant to the current session's nodes.

\section{Conclusions}

We have proposed a novel method I3GN to integrate inter-session item-level interactions into session-based recommendations. I3GN is consist of two major parts: SMM and CSE. SMM uses similarity to find neighbor sessions and constructs the intra-session graph and inter-session graph for the current session based on them. After that, CSE employs different strategies to encode two kinds of graphs and combine them to get the final session representation for prediction. Extensive experiments on real-world datasets prove that I3GN outperforms other state-of-the-art methods in different evaluation metrics. Further experiments and analysis demonstrate the following facts: (1) Inter-session item-level interactions have high potential in session-based recommendations. (2) Combining intra-and inter-session graphs is a rational way to balance the model across scopes, and its superior performance indicates that we need to take both the width and depth of session data into account.

\bibliographystyle{ACM-Reference-Format}
\bibliography{sample-base}

\end{document}